\documentclass[twocolumn,showpacs,preprintnumbers,amsmath,amssymb,floatfix,a4paper]{revtex4}

\usepackage{graphicx}
\usepackage{dcolumn}
\usepackage{bm}

\begin{document}

\title{Examination of the role of the $^{14}$O($\alpha$,$p$)$^{17}$F reaction rate in type I x-ray bursts}

\author{J.~Hu$^{1,2}$}
\author{J.J.~He$^1$}
\email{jianjunhe@impcas.ac.cn}
\author{A.~Parikh$^{3,4}$}
\email{anuj.r.parikh@upc.edu}
\author{S.W.~Xu$^{1,5}$}
\author{H.~Yamaguchi$^2$}
\author{D.~Kahl$^{2}$}
\author{P.~Ma$^{1}$}
\author{J.~Su$^{6}$}
\author{H.W.~Wang$^{7}$}
\author{T.~Nakao$^{2}$}
\author{Y.~Wakabayashi$^{8}$}
\author{T.~Teranishi$^{9}$}
\author{K.I.~Hahn$^{10}$}
\author{J.Y.~Moon$^{11}$}
\author{H.S.~Jung$^{11}$}
\altaffiliation[Present address: ]{Department of Physics, University of Notre Dame, Indiana 46556, USA.}
\author{T.~Hashimoto$^{12}$}
\author{A.A.~Chen$^{13}$}
\author{D.~Irvine$^{13}$}
\author{C.S.~Lee$^{11}$}
\author{S.~Kubono$^{1,8}$}

\affiliation{$^1$Key Laboratory of High Precision Nuclear Spectroscopy and Center for Nuclear Matter Science, Institute of Modern Physics, Chinese Academy of Sciences, Lanzhou 730000, China}
\affiliation{$^2$Center for Nuclear Study (CNS), the University of Tokyo, Wako Branch at RIKEN, 2-1 Hirosawa, Wako, Saitama 351-0198, Japan}
\affiliation{$^3$Departament de F\'{\i}sica i Enginyeria Nuclear, EUETIB, Universitat Polit\`{e}cnica de Catalunya, Barcelona E-08036, Spain}
\affiliation{$^4$Institut d'Estudis Espacials de Catalunya, Barcelona E-08034, Spain}
\affiliation{$^5$University of Chinese Academy of Sciences, Beijing 100049, China}
\affiliation{$^6$China Institute of Atomic Energy (CIAE), P.O. Box 275(46), Beijing 102413, China}
\affiliation{$^7$Shanghai Institute of Applied Physics (SINAP), Chinese Academy of Sciences (CAS), Shanghai 201800, China}
\affiliation{$^8$RIKEN Nishina Center, 2-1 Hirosawa, Wako, Saitama 351-0198, Japan}
\affiliation{$^9$Department of Physics, Kyushu University, 6-10-1 Hakozaki, Fukuoka 812-8581, Japan}
\affiliation{$^{10}$Department of Science Education, Ewha Womans University, Seoul 120-750, Republic of Korea}
\affiliation{$^{11}$Department of Physics, Chung-Ang University, Seoul 156-756, Republic of Korea}
\affiliation{$^{12}$Research Center for Nuclear Physics (RCNP), Osaka University, 10-1 Mihogaoka, Ibaraki, Osaka, 567-0047, Japan}
\affiliation{$^{13}$Department of Physics \& Astronomy, McMaster University, Hamilton, Ontario L8S 4M1, Canada}


\begin{abstract}
The $^{14}$O($\alpha$,$p$)$^{17}$F reaction is one of the key reactions involved in the breakout from the hot-CNO cycle to the rp-process
in type I x-ray bursts (XRBs). The resonant properties in the compound nucleus $^{18}$Ne have been investigated through resonant elastic
scattering of $^{17}$F+$p$. The radioactive $^{17}$F beam was separated by the CNS Radioactive Ion Beam separator (CRIB) and bombarded a
thick H$_2$ gas target at 3.6 MeV/nucleon. The recoiling light particles were measured by three ${\Delta}$E-E silicon telescopes at
laboratory angles of $\theta$$_{lab}$$\approx$3$^\circ$, 10$^\circ$ and 18$^\circ$, respectively. Five resonances at $E_{x}$=6.15, 6.28,
6.35, 6.85, and 7.05 MeV were observed in the excitation functions, and their spin-parities have been determined based on an $R$-matrix
analysis. In particular, $J^{\pi}$=1$^-$ was firmly assigned to the 6.15-MeV state which dominates the thermonuclear
$^{14}$O($\alpha$,$p$)$^{17}$F rate below 2 GK. As well, a possible new excited state in $^{18}$Ne was observed at $E_{x}$=6.85$\pm$0.11 MeV
with tentative $J$=0 assignment. This state could be the analog state of the 6.880 MeV (0$^{-}$) level in the mirror nucleus $^{18}$O, or
a bandhead state (0$^+$) of the six-particle four-hole (6$p$-4$h$) band. A new thermonuclear $^{14}$O($\alpha$,$p$)$^{17}$F rate has been
determined, and the astrophysical impact of multiple recent rates has been examined using an XRB model. Contrary to
previous expectations, we find only modest impact on predicted nuclear energy generation rates from using reaction rates differing by up
to several orders of magnitude.
\end{abstract}



\maketitle

Type I x-ray bursts (XRBs) are characterized by sudden dramatic increases in luminosity of roughly 10--100 s in duration, with peak
luminosities of roughly 10$^{38}$ erg/s. These recurrent phenomena (on timescales of hours to days) have been the subject of many
observational, theoretical and experimental studies (for reviews see {\it e.g.},~\cite{bib:lew93,bib:str06,bib:par13}). The bursts have
been interpreted as being generated by thermonuclear runaway on the surface of a neutron star that accretes H- and He-rich material from
a less evolved companion star in a close binary system~\cite{bib:woo76,bib:jos77}. The accreted material burns stably through the hot,
$\beta$-limited carbon-nitrogen-oxygen (HCNO)~\cite{bib:wie99} cycles, giving rise to the persistent flux. Once critical temperatures
and densities are achieved, breakout from this region toward higher masses can occur through alpha-induced reactions. Subsequently, the
rapid-proton capture (rp) process drives nucleosynthesis toward the proton drip-line~\cite{bib:wal81,bib:sch98,bib:woo04}. This
eventually results in a rapid increase in energy generation (ultimately leading to the XRB) and nucleosynthesis up to A$\sim$100 mass
region~\cite{bib:sch01,bib:elo09}.

It has long been known that helium burning on HCNO seeds ({\it e.g.}, $^{14}$O and $^{15}$O) drives the thermonuclear runaway and that
the $^{14}$O($\alpha$,$p$)$^{17}$F reaction initiates one of the reaction sequences leading to breakout from the HCNO
cycles~\cite{bib:wie99,bib:wal81,bib:taa85,bib:woo84,bib:wie98,bib:taa79}. The astrophysical impact of different calculated
$^{14}$O($\alpha$,$p$)$^{17}$F rates (as well as associated uncertainties) has not, however, been carefully assessed. The precise rate
of this reaction has previously been suspected to be of only secondary importance~\cite{bib:fuj81,bib:ili07} yet the need for improved
determinations was nonetheless repeatedly stressed~\cite{bib:par13,bib:wie99,bib:sch99,bib:wie87,bib:bla03,bib:sch06,bib:par08}.
In addition, variations of an adopted rate by constant factors in different one-zone XRB models has had possibly inconsistent effects
on predicted energy generation rates~\cite{bib:par08,bib:amt08}. As such, it is of interest not only to resolve discrepancies in recent
$^{14}$O($\alpha$,$p$)$^{17}$F rate calculations through new measurements but also to actually evaluate the astrophysical impact of recent
rates using a consistent set of XRB model calculations.

Contributions from $^{14}$O+$\alpha$ resonances in $^{18}$Ne ($Q_\alpha$=5.115 MeV~\cite{bib:wan12}) dominate the $^{14}$O($\alpha$,$p$)$^{17}$F
rate at $T$ relevant to XRBs ($\approx$0.2--2 GK). Although our understanding of this rate has been improved by indirect
studies~\cite{bib:bla03,bib:hah96,bib:par99,bib:park,bib:gom01,bib:cern,bib:bar10,bib:he11}, direct study~\cite{bib:not04}, as well as
time-reversal studies~\cite{bib:bla01,bib:har99,bib:har02}, most of the required resonance parameters (such as , $J^{\pi}$ and $\Gamma_\alpha$)
have still not been sufficiently well determined.

In the temperature region below $\sim$1 GK, a state at $E_x$=6.15 MeV (tentatively assigned as 1$^-$, see below) was thought to dominate
the $^{14}$O($\alpha$,$p$)$^{17}$F rate~\cite{bib:hah96}. About twenty-five years ago, Wiescher {\it et al.}~\cite{bib:wie87} predicted
a $J^\pi$=1$^-$ state at $E_x$=6.125 MeV in $^{18}$Ne with a width of $\Gamma$=$\Gamma_p$=51 keV based on a Thomas-Ehrman shift calculation.
Later on, Hahn {\it et al.}~\cite{bib:hah96} observed a state at $E_x$=6.15$\pm$0.01 MeV through studies of the $^{16}$O($^3$He,$n$)$^{18}$Ne
and $^{12}$C($^{12}$C,$^6$He)$^{18}$Ne reactions. The transferred angular momentum was restricted to $\ell$$\leq$2 from the ($^3$He,$n$)
angular distribution measured. Based on the Coulomb-shift calculation and prediction of Wiescher {\it et al.}, a $J^\pi$=1$^-$ was tentatively
assigned to this state. G\"{o}mez {\it et al.}~\cite{bib:gom01} studied the resonances in $^{18}$Ne by using the elastic scattering of
$^{17}$F+$p$ and fitted the 6.15-MeV state with 1$^-$ by an $R$-matrix code. However, their 1$^{-}$ assignment was questioned in a later
$R$-matrix reanalysis~\cite{bib:arX}. He {\it et al.}~\cite{bib:arX} thought that this 1$^-$ resonance should behave as a dip-like structure
(rather than the peak observed in Ref.~\cite{bib:gom01}) in the excitation function due to the interference. Unfortunately, our previous
low-statistics measurement could not resolve this state~\cite{bib:he11}. Recently, Bardayan {\it et al.}~\cite{bib:bar12} reanalyzed the
unpublished elastic-scattering data in Ref.~\cite{bib:bla03} and identified the expected dip-like structure, however, the statistics were
not sufficient to constrain the parameters of such a resonance. Therefore, three possibilities arise regarding the results of
Ref.~\cite{bib:gom01} on the $J^\pi$ of the 6.15 MeV state:
(i) their analysis procedure may be questionable as they reconstructed the excitation functions (above 2.1 MeV) with some technical treatment
since the high-energy protons escaped from two thin Si detectors;
(ii) the peak observed in Ref.~\cite{bib:gom01} may be due to the inelastic scattering contribution~\cite{bib:bar12,bib:gri02}, or the
carbon-induced background (from CH$_2$ target itself) which was not measured and subtracted accordingly;
(iii) the 1$^-$ assignment for the 6.15-MeV state was wrong in Ref.~\cite{bib:gom01}. If their data were correct, the results~\cite{bib:arX}
show that the 6.15-MeV state most probably has a 3$^-$ or 2$^-$ assignment, while the 6.30-MeV state becomes the key 1$^-$ state. In addition,
the inelastic branches of $^{17}$F($p$,$p^\prime$)$^{17}$F$^\ast$ (not measured in Ref.~\cite{bib:gom01}) can contribute to the
$^{14}$O($\alpha$,$p$)$^{17}$F reaction rate considerably. Constraining the proton-decay branches to the ground and first excited ($E_x$=495 keV,
$J^\pi$=1/2$^+$) states of $^{17}$F is therefore of critical importance. Such inelastic channels were observed for several $^{18}$Ne
levels~\cite{bib:bla03,bib:cern,bib:not04,bib:bar12,bib:alm12}, however, there are still some controversies~\cite{bib:for12}.

In this work, we first address outstanding uncertainties of relevant $^{14}$O+$\alpha$ resonances in $^{18}$Ne through a new $^{17}$F+$p$
resonant elastic scattering measurement in inverse kinematics, with a $^{17}$F radioactive ion beam. The thick target
method~\cite{bib:dae64,bib:art90,bib:gal91,bib:axe96,bib:kub01}, proven to be successful in our previous
studies~\cite{bib:ter03,bib:hjj07,bib:yam09,bib:korea}, was used in this experiment. In particular, we have unambiguously determined the 1$^-$
character of the important 6.15 MeV resonance. We then use all available experimental input to calculate a new $^{14}$O($\alpha$,$p$)$^{17}$F
rate at temperatures involved in XRBs. Finally, we have examined the impact of multiple recent $^{14}$O($\alpha$,$p$)$^{17}$F rates within the
framework of one-zone XRB postprocessing calculations.

The experiment was performed using the CNS Radioactive Ion Beam separator (CRIB)~\cite{bib:yan05,bib:kub02}, installed by the Center for
Nuclear Study (CNS), the University of Tokyo, in the RI Beam Factory of RIKEN Nishina Center. A primary beam of $^{16}$O$^{6+}$ was accelerated
up to 6.6 MeV/nucleon by an AVF cyclotron ($K$=80) with an average intensity of 560 enA. The primary beam delivered to CRIB bombarded a
liquid-nitrogen-cooled D$_{2}$ gas target ($\sim$90 K)~\cite{bib:yam08} where $^{17}$F RI beam was produced via the $^{16}$O($d$,$n$)$^{17}$F
reaction in inverse kinematics. The D$_{2}$ gas at 120 Torr pressure was confined in a 80-mm long cell with two 2.5 ${\mu}$m thick Havar foils.
The $^{17}$F beam was separated by the CRIB. The $^{17}$F beam, with a mean energy of 61.9$\pm$0.5 MeV (measured by a silicon detector) and
an average intensity of 2.5${\times}$10$^{5}$ pps, bombarded a thick H$_{2}$ gas target in a scattering chamber located at the final focal
plane (F3); the beam was stopped completely in this target.

\begin{figure}
\begin{center}
\includegraphics[width=8.6cm]{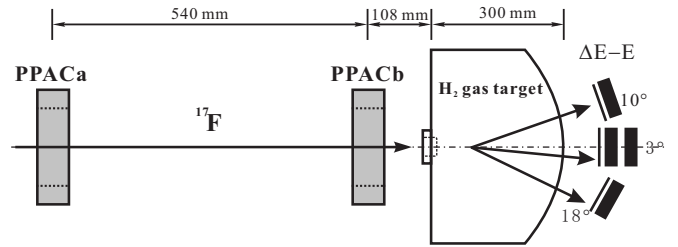}
\end{center}
\vspace{-6mm}
\caption{\label{fig1} Schematic diagram of the experimental setup at the scattering chamber, similar to that used in Ref.~\cite{bib:korea}.}
\end{figure}

\begin{figure}
\includegraphics[width=8cm]{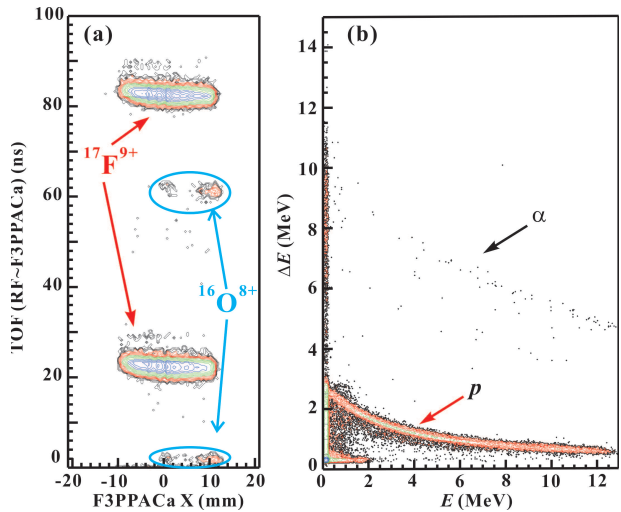}
\vspace{-3mm}
\caption{\label{fig2} (Color online) (a) Identification plot for the beam particles before H$_2$ target via time-of-flight (TOF) technique.
Two groups of particles appear for a single beam, since the data for two extraction cycles of the cyclotron are plotted together.
(b) Identification plot for the recoiled particles via $\Delta E-E$ technique. See text for details.}
\end{figure}

The experimental setup at the F3 chamber is shown in Fig.~\ref{fig1}, which is quite similar to that used in Ref.~\cite{bib:korea}.
The beam purity was about 98\% after the Wien-filter. Two PPACs (Parallel Plate Avalanche Counters)~\cite{bib:kum01} provided the timing
and two-dimensional position information of the beam particles. The beam profile on the secondary target was monitored by the PPACs during
the data acquisition. The beam particles were identified event-by-event by the time of flight (TOF) between PPACa (see Fig.~\ref{fig1}) and
the production target using the RF signal provided by the cyclotron. Figure~\ref{fig2}(a) shows the particle identification at PPACa. The
H$_{2}$ gas target at a pressure of 600 Torr was housed in a 300-mm-radius semi-cylindrical chamber sealed with a 2.5-$\mu$m-thick Havar
foil as an entrance window and a 25-$\mu$m-thick aluminized Mylar foil as an exit window. Comparing to the previously used solid CH$_{2}$
target~\cite{bib:bla03,bib:gom01,bib:cern,bib:he11,bib:har99,bib:har02}, the gas target is free from intrinsic background from carbon.

The recoiling light particles were measured by three ${\Delta}$E-E Si telescopes at average angles of $\theta$$_{lab}$$\approx$3$^\circ$,
10$^\circ$ and 18$^\circ$, respectively. In the {\it c.m.} frame of elastic scattering, the corresponding scattering angles are
$\theta_{c.m.}$$\approx$155$^\circ$$\pm$18$^\circ$, 138$^\circ$$\pm$22$^\circ$ and 120$^\circ$$\pm$22$^\circ$, respectively. At
$\theta$$_{lab}$$\approx$3$^\circ$, the telescope consisted of a 65-$\mu$m-thick double-sided-strip (16$\times$16 strips) silicon detector
and two 1500-$\mu$m-thick pad detectors. The last pad detector was used to veto any energetic light ions produced in the production target
and satisfying the $B\rho$ selection, but not rejected by the Wien filter because of scattering in the inner wall of the beam line. The
configuration of the other two telescopes is similar to that at $\theta$$_{lab}$$\approx$3$^\circ$, except for the absence of the third
veto layer. The position sensitive $\Delta$E detectors measured the energy, position and timing signals of the particles, and the pad E
detectors measured their residual energies. The recoiling particles were clearly identified by using a $\Delta E-E$ method as shown in
Fig.~\ref{fig2}(b). The energy calibration for the silicon detectors was performed by using a standard triple ${\alpha}$ source and
secondary proton beams at several energy points produced with CRIB during calibration runs. The contribution of background was evaluated
through a separate run with Ar gas at 120 Torr (chosen to achieve the equivalent stopping power as in the H$_{2}$ gas).

\begin{figure}
\includegraphics[width=8.6cm]{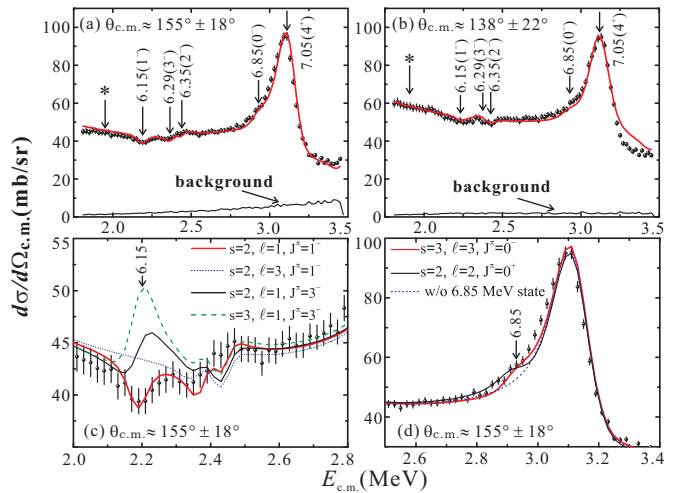}
\vspace{-6mm}
\caption{\label{fig3} (Color online) The center-of-mass differential cross-sections for elastically scattered protons of $^{17}$F+$p$ at
angles of (a)$\theta$$_{c.m.}$$\approx$155$^\circ$$\pm$18$^\circ$, and (b)$\theta$$_{c.m.}$$\approx$138$^\circ$$\pm$22$^\circ$. The (red)
curved lines represent the best overall $R$-matrix fits. The locations of inelastic scattering events for the 6.15-MeV state are indicated
as the asterisks. The indicated background spectra (from the Ar gas run) was subtracted accordingly. Additional $R$-matrix fits for the
6.15- and 6.85-MeV states are shown in (c) and (d), respectively. See text for details.}
\end{figure}

The excitation functions of $^{17}$F+$p$ elastic scattering have been reconstructed using the procedure described
previously~\cite{bib:he11,bib:kub01,bib:hjj07}. The excitation functions at two scattering angles are shown in Fig.~\ref{fig3}; data from
the third telescope (at $\theta$$_{lab}$$\approx$18$^\circ$) were not included in the analysis due to its considerably poorer resolution.
The normalized Ar-gas background spectra shown was subtracted accordingly. Our results demonstrate that the pure H$_{2}$ gas target allows
us to minimize the background protons. It can be regarded as a strong merit compared to the generally used CH$_2$ solid target which
contributes significantly more background from C atoms. The length of the gas target (300 mm) led to an uncertainty of about 3\% in the
solid angle, as determined in event-by-event mode. Such uncertainty in the cross-section is comparable to the statistical one ($\approx$1\%).

Several resonant structures were clearly observed in the spectra. In order to determine the resonant parameters of observed resonances,
multichannel $R$-matrix calculations~\cite{bib:lan58,bib:des03,bib:bru02} (see examples~\cite{bib:arX,bib:mur09}) have been performed in
the present work. A channel radius of $R$=1.25$\times$(1+17$^{\frac{1}{3}}$)$\approx$4.46 fm appropriate for the $^{17}$F+$p$
system~\cite{bib:wie87,bib:hah96,bib:gom01,bib:he11,bib:arX,bib:nel85} has been utilized in the calculation. The choice of radius only
has minor effect on the large uncertainties quoted both for the excitation energies and widths (see Table~\ref{table1}).

The ground-state spin-parity configurations of $^{17}$F and proton are $5/2^{+}$ and $1/2^{+}$, respectively. Thus, there are two channel
spins in the elastic channel, {\it i.e.}, $s$=2 and 3. In the present $R$-matrix calculation, the $\alpha$ partial widths ($\Gamma_\alpha$)
are negligible relative to the proton widths ($\Gamma_\alpha$$\ll$$\Gamma_p$)~\cite{bib:hah96,bib:har02}. Five resonances, at $E_{x}$=6.15,
6.28, 6.35, 6.85, and 7.05 MeV, have been analyzed, and the best overall fitting curves are shown in Fig.~\ref{fig3}(a)\&(b). The resonant
parameters obtained are listed in Table~\ref{table1}. In order to fit the data around $E_{c.m.}$=3.2 MeV, it was necessary to include an
additional known resonance ($E_x$$\sim$7.40 MeV, $J^\pi$=2$^+$, $\Gamma$=40 keV)~\cite{bib:he11,bib:har02,bib:for00} in the calculations.

\vspace{2mm}
\noindent\emph{(a) States between 6.1--6.4 MeV}

According to the $R$-matrix analysis, a dip-like structure around $E_{c.m.}$=2.21 MeV, corresponding to the 6.15-MeV state in $^{18}$Ne,
is best fit as a natural-parity 1$^-$ state as shown in Fig.~\ref{fig3}(c). The fitted parameters are $\ell$=1, $s$=2, and $\Gamma$=50$\pm$15 keV
(see Table~\ref{table1}), where the measured total width $\Gamma$ is consistent with $\Gamma$=53.7$\pm$2.6 keV reported before~\cite{bib:bar12}.
The resonance shape of this state agrees with that of a previous low-statistics experiment~\cite{bib:bla03} reanalyzed by Bardayan
{\it et al.}~\cite{bib:bar12}. The natural-parity character of this state was verified by the previous direct $^{14}$O($\alpha$,$p$)$^{17}$F
measurement~\cite{bib:not04}. Thus, the unnatural-parity 2$^-$ assignment can be excluded, and such assignment is also unlikely based on the
discussions of the $2p$-emission from this state~\cite{bib:gom01,bib:rac08}. In addition, the 3$^-$ assignment is very unlikely as shown in
Fig.~\ref{fig3}(c), and such assignment can also be ruled out because of the large inelastic branch observed for this state. Therefore, we
confirmed the 1$^-$ assignment of the important 6.15-MeV state. Our resonance shape is entirely different from the bump-like shape observed in
Ref.~\cite{bib:gom01}. This may be due to issues in the data of Ref.~\cite{bib:gom01} as well as their $R$-matrix analysis (see the lower panel
of Fig.~2 in Ref.~\cite{bib:gom01}). As a result, other $J^\pi$ assignments suggested in Ref.~\cite{bib:arX} are also questionable.

A structure at $E_{x}$=6.28 MeV was observed in the excitation function, and its shape is reproduced with those resonant parameters from the
work of Hahn {\it et al.}, {\it i.e.}, $E_{c.m.}$=2.36 MeV, $J^\pi$=3$^-$, and $\Gamma$=20 keV. In Ref.~\cite{bib:gom01}, this state was
not involved in their $R$-matrix fit. This natural-parity state was clearly observed in the direct $^{14}$O($\alpha$,$p$)$^{17}$F
experiment~\cite{bib:not04}.

The 6.35-MeV state is fitted well with parameters of $J^\pi$=2$^-$, and $\Gamma$=10$\pm$5 keV. This $J^\pi$ assignment is consistent with
that speculated by Hahn {\it et al.} It was only weakly populated in the transfer reactions of ($^3$He,$n$) and ($p$,$t$), and unobserved
in the direct $^{14}$O($\alpha$,$p$)$^{17}$F experiment~\cite{bib:not04}. With an unnatural-parity 2$^-$ assignment, this state does not
contribute to the rate~\cite{bib:hah96,bib:har02}.

In summary up to this point, we have made confirmation of the three states between 6.1 and 6.4 MeV for the first time, which has been a long
standing problem~\cite{bib:hah96,bib:park}. Because of nuclear structure considerations (4$p$-2$h$ configuration of $h$ (hole) being in $1p$3/2
and $p$ (particle) in $2s$1/2 or $1d$3/2 orbits), 1$^-$ has very small ($p$,$t$) cross section, and that is why the 6.15-MeV state was not
observed in the previous experiments~\cite{bib:hah96,bib:park}. On the other hand, the 2$^-$ state can be expected to have appreciable amplitude
with a simple $p$-$h$ component, since there is always ($p$,$t$) multistep component even for an unnatural-parity state~\cite{bib:hah96}. That
is why the 6.35-MeV state could be observed even by the ($p$,$t$) reactions~\cite{bib:hah96,bib:park}; but this 2$^-$ amplitude is significantly
smaller than that of 3$^-$ natural-parity state at 6.286-MeV.

The first study to observe inelastic scattering from the 6.15-MeV state was reported by Blackmon {\it et al.}~\cite{bib:bla03}. They yielded
a branching ratio of $\Gamma_{p\prime}$/$\Gamma_{p}$=2.4, and total $\Gamma$$\sim$58 keV, where $\Gamma_{p}$ and $\Gamma_{p\prime}$ are
the proton-branching widths for populating the ground and first excited states, respectively. He {\it et al.}~\cite{bib:cern} detected decay
$\gamma$ rays in coincidence with $^{17}$F+$p$ protons looking at the 495-keV $\gamma$ rays, and yielded a ratio of
$\Gamma_{p\prime}$/$\Gamma_{p}$$\sim$1. By reanalyzing the data in Ref.~\cite{bib:bla03}, Bardayan {\it et al.}~\cite{bib:bar12} derived a
new ratio of $\Gamma_{p\prime}$/$\Gamma_{p}$=0.42$\pm$0.03, and $\Gamma$=53.7$\pm$2.0 keV. Most recently, Almaraz-Calderon
{\it et al.}~\cite{bib:alm12} populated the 6.15-MeV state via the $^{16}$O($^3$He,$n$)$^{18}$Ne reaction. Due to large uncertainties,
they only estimated the upper limit of this branching ratio ($\Gamma_{p\prime}$/$\Gamma_{p}$$\leq$0.27). Furthermore, the resolution in
the TOF spectrum could result in a relatively large uncertainty in the excitation energies (see Figure 6 in Ref.~\cite{bib:alm12}).
In Fig.~\ref{fig3}(a)\&(b), the position of the inelastic scattering events is indicated for the 6.15-MeV state. However, no noticeable structure
was observed because of the smaller amplitude for this inelastic channel, {\it i.e.}, less than half that of the elastic one.
The inelastic channel was not included in the present $R$-matrix analysis, where inclusion of a small non-zero $\Gamma_{p\prime}$ has no
effect on the conclusion regarding $J^\pi$=1$^-$ discussed above.

A shell-model calculation for A=17 and 18 nuclides has been performed with a shell-model code OXBASH~\cite{bib:bro92}. The calculation was
carried out in a full model space (spsdpf) using an isospin-conserving WBB interaction of Warburton and Brown~\cite{bib:war92}. The energy of
the second 1$^-$ state was predicted to be $E_x$=6.652 MeV for $^{18}$Ne and $^{18}$O. According to the knowledge of the mirror
$^{18}$O~\cite{bib:li76}, this 1$^-$ state originates mainly from the valence hole of $1p_{3/2}$. The spectroscopic factors are calculated to
be about $S_p(1p_{3/2})$=0.01 for both proton decays to the ground and the first excited states in $^{17}$F. The calculated value of $S$ is
about three times smaller than the experimental one~\cite{bib:li76} in $^{18}$O. Due the complicated configuration mixing, the theoretical
value may fail to reproduce the absolute experimental $S$ value, but the spectroscopic factor ratio between the ground and first excited state
should be reliable. The calculated branching ratio is $\Gamma_{p\prime}$/$\Gamma_{p}$$\approx$0.66 with a partial proton width relation of
$\Gamma_{p}$=$\frac{3\hbar^2}{\mu R^2}$$P_{\ell}$$C^2S_{p}$~\cite{bib:wie87}. The calculated proton width is about 20 keV with $C^2S_{p}$=0.01.
These results are reasonable given the reanalysis by Bardayan {\it et al.}~\cite{bib:bar12}

\vspace{2mm}
\noindent\emph{(b) State at 6.85 MeV}

It is very interesting that a shoulder-like structure around $E_{c.m.}$=2.93 MeV was observed by both telescopes as shown in Fig.~\ref{fig3}(a)\&(b).
This is possibly a new state at $E_{x}$=6.85$\pm$0.10 MeV. Both $J^\pi$=0$^-$ or 0$^+$ resonances can reproduce the observed shape as shown
in Fig.~\ref{fig3}(d). Because of the small energy shift for the negative-parity states in this excitation energy region~\cite{bib:for00},
such a state is possibly the analog state of $^{18}$O at $E_x$=6.880 MeV (0$^-$)~\cite{bib:til95}. In fact, Wiescher {\it et al.}~\cite{bib:wie87}
predicted a $J^\pi$=0$^-$ state in $^{18}$Ne, analog to the 6.88 MeV state in $^{18}$O, at 6.85 MeV with a proton spectroscopic factor of
$C^2S_p$=0.01. However, another possibility still exists.

A strong proton resonance from a state at $E_x$$\sim$6.6 MeV was observed in an earlier direct $^{14}$O($\alpha$,$p$)$^{17}$F
experiment~\cite{bib:not04}. Because no such state was previously observed in $^{18}$Ne, Notani {\it et al.} speculated that it might be due to a
state at $E_x$$\sim$7.1 MeV decaying to the first excited state of $^{17}$F. Later on, a careful $^{17}$F+$p$ scattering experiment~\cite{bib:bar10}
was performed, but no evidence of inelastic $^{17}$F+$p$ scattering was observed in this energy region, and the decay branching ratio to the
first excited state ($\Gamma_{p\prime}$/$\Gamma_{p}$) was constrained to be $<$0.03. Almaraz-Calderon {\it et al.} recently reported a ratio of
0.19$\pm$0.08 for the 7.05 MeV state. Later on, this large ratio was questioned by Fortune~\cite{bib:for12} who estimated a ratio less than
about 2$\times$10$^{-4}$, in agreement with an earlier limit of $\leq$1/90 from Harss {\it et al.}~\cite{bib:har02}. Based on the suggestion
of Fortune, Almaraz-Calderon {\it et al.} thought that their large number might be attributed to an unknown state at $E_x$$\sim$6.7 MeV in
$^{18}$Ne. In fact, there is a hint of a weak state observed at $E_x$$\sim$6.8 MeV (see Figure 6 in Ref.~\cite{bib:alm12}). As discussed above,
such a state at $E_x$=6.85$\pm$0.10 MeV was also observed in the present work. Therefore, we conclude that very likely a new state around
6.8 MeV exists in $^{18}$Ne. Since this state was populated in the direct $^{14}$O($\alpha$,$p$)$^{17}$F reaction, it should have a natural parity.
Thus, it is also possibly a candidate for the $J^\pi$=0$^+$ state, a bandhead state of the six-particle four-hole (6$p$-4$h$)
band~\cite{bib:for03,bib:for11}. If this 6.85-MeV state were 0$^+$, its $\alpha$ width would be roughly 149 eV, as estimated with the expression
of $\Gamma_{\alpha}$=$\frac{3\hbar^2}{\mu R^2}$$P_{\ell}(E)$$C^2S_\alpha$~\cite{bib:wie87}. Here, a spectroscopic factor of 0.01 was assumed in
the calculation. As such, if the state is 0$^+$ ($\omega\gamma$=149 eV), its contribution to the $^{14}$O($\alpha$,$p$)$^{17}$F rate would be
larger than that of the 7.05-MeV state ($\omega\gamma$=203 eV) but still much smaller than that of the 6.15 MeV state below $\sim$2.5 GK. Of course,
if it is, in fact, 0$^-$, it would not contribute to the rate at all. The exact $J^\pi$ for this 6.85 MeV state still needs to be determined by
additional experiments (though we prefer 0$^+$).

\vspace{2mm}
\noindent\emph{(c) States at 7.05 and 7.35 MeV}

In this work, a state at $E_{x}$=(7.05$\pm$0.03) MeV (4$^+$, $\Gamma$=95 keV)~\cite{bib:har02} was observed at $E_{c.m.}$=3.13 MeV. However, the
doublet structure around $E_{x}$=7.05 and 7.12 MeV suggested in Refs.~\cite{bib:hah96,bib:he11} could not be resolved within the present energy
resolution ($\sim$80 keV in FWHM in this region). A single peak is adequate for the fit to our data, with similar $\chi$$^2$ value to a fit using two peaks.

One state at (7.35$\pm$0.02) MeV was observed in the ($^3$He,$n$) and ($^{12}$C,$^6$He) reactions~\cite{bib:hah96} and showed (1$^-$, 2$^+$)
characteristics in the ($^3$He,$n$) angular distribution. Hahn {\em et al.}~\cite{bib:hah96} suggested a 1$^-$ for this state based on a very
simple mirror argument. Later on, following the arguments of Fortune and Sherr~\cite{bib:for00}, Harss {\em et al.}~\cite{bib:har02} speculated
it, {\it i.e.}, at (7.37$\pm$0.06) MeV, as a 2$^+$ state based on a Coulomb-shift discussion. Our present and previous results~\cite{bib:he11}
all support the 2$^+$ assignment. However, its mirror partner is still uncertain~\cite{bib:for11}. Combining with the discussion of Fortune and
Sherr~\cite{bib:for11}, we speculate that a new 7.796-MeV state recently observed~\cite{bib:oer10} in $^{18}$O may be the mirror of the 7.35 MeV
state in $^{18}$Ne. This would imply that the bandhead (0$^+$) of the six-particle four-hole (6$p$-4$h$)~\cite{bib:for03,bib:for11} band in
$^{18}$O is still missing.

\begin{figure}
\includegraphics[width=8.6cm]{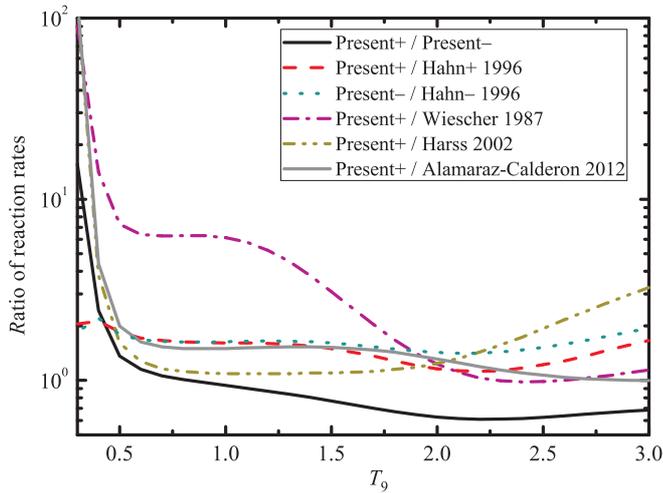}
\vspace{-6mm}
\caption{\label{fig4} (Color online) Ratios between the present reaction rates and previous calculations~\cite{bib:hah96,bib:alm12,bib:wie87,bib:har02}.
See text for details.}
\end{figure}

By evaluating all the available data, the resonance parameters adopted for the $^{14}$O($\alpha$,$p$)$^{17}$F resonant rate calculations are
summarized in Table~\ref{table2}. Here, the excitation and resonance energies are adopted from the work of Hahn {\it et al.}~\cite{bib:hah96}.
Similar to the method utilized by Hahn {\it et al.} and Bardayan {\it et al.}~\cite{bib:bar97}, the $^{14}$O($\alpha$,$p$)$^{17}$F total rate
has been numerically calculated using the resonance parameters listed in Table~\ref{table2} and the direct reaction $S$-factors calculated by
Funck \& Langanke~\cite{bib:fun88}. Here, the interference between the direct-reaction $\ell$=1 partial wave and the 6.15-MeV (1$^{-}$) excited
state was included in the calculations; the inelastic branches (listed in Table~\ref{table2}) were also included in the integration. Two different
$^{14}$O($\alpha$,$p$)$^{17}$F rates were calculated by assuming the constructive (``Present+") and destructive (``Present-") interferences
between the direct and resonant captures (for the 6.15-MeV state). These two rates differ by a factor of $\approx$5 at 0.35 GK and less than
10\% at 1 GK. In the temperature region of 0.3--3 GK, our ``Present+" and ``Present-" rates are about 1.1--2.2 times larger than the corresponding
rates from Hahn {\it et al.}. Our adopted parameters are more reliable than the older ones determined by Hahn {\it et al.} about twenty years ago.
It is worth noting that our rates are orders of magnitude greater than those of Harss {\it et al.}~\cite{bib:har02} and Alamaraz-Calderon
{\it et al.}~\cite{bib:alm12} below 0.3 GK, because they did not consider the interference effects, and only utilized a simple narrow-resonance
formalism to calculate the resonant rate of the 6.15-MeV state. The comparison between our rates and the previous ones are shown in Fig.~\ref{fig4}.
The 1$\sigma$ uncertainties (lower and upper limits as utilized below) of the present rates were estimated to be about 10--30\%
(for ``Present+") and 20--50\% (for ``Present-") over 0.1--3 GK, using a Monte-Carlo method with parameter errors adopted in Table~\ref{table2}.
We found that the contribution from the 6.15 MeV state dominates the total rate over temperatures of interest in XRBs. The present (+/-) recommended
rates for $T_9$$\le$3 GK can be analytically expressed by
$N_{A}\langle\sigma v\rangle^+_{-}$=$\mathrm{exp} [^{1.890}_{1.301}\times 10^2 - ^{0.179}_{0.157} T_9^{-1} - ^{4.872}_{10.731} T_9^{-1/3}
 - ^{2.358}_{1.445}\times 10^2 T_9^{1/3} + ^{74.11}_{18.51} T_9 - ^{22.61}_{1.67} T_9^{5/3}
+ ^{52.71}_{38.83} \mathrm{ln}T_9] + \mathrm{exp} [- ^{2.149}_{1.677}\times 10^2 + ^{11.843}_{10.333} T_9^{-1}
 - ^{9.092}_{7.892}\times 10^2 T_9^{-1/3}
 - ^{1.166}_{0.994}\times 10^3 T_9^{1/3} - ^{56.26}_{49.47} T_9 + ^{26.82}_{24.73} T_9^{5/3} - ^{6.179}_{5.277}\times 10^2 \mathrm{ln}T_9]$.

\begin{table}
\caption{\label{table1} Resonant parameters derived from the present $R$-matrix analysis. The excitation energies are the average values derived
from our data sets, and uncertainties are estimated by a Monte-Carlo simulation. The widths available in the literature are listed for comparison.}
\begin{tabular}{|l|c|c|c|l|}
\hline \hline
$E_x$ (MeV) & $J^\pi$ & $\ell$ & $\Gamma$ (keV)$^\mathrm{present}$ & $\Gamma$ (keV)$^\mathrm{literature}$                 \\
\hline
6.15(0.03)     & 1$^{-}$ & 1  &  50(15)  & $\leq$40~\cite{bib:hah96}; 53.7$\pm$2.0~\cite{bib:bar12} \\
6.28(0.03)     & 3$^{-}$ & 1  &  20(15)  & $\leq$20~\cite{bib:hah96}; 8$\pm$7~\cite{bib:par99}      \\
6.35(0.03)     & 2$^{-}$ & 1  &  10(5)   & 45$\pm$10~\cite{bib:hah96}; 18$\pm$9~\cite{bib:par99}    \\
6.85(0.11)     & 0$^{-}$ & 3  &  50(30)  &  \\
               & 0$^{+}$ & 2  &  50(30)  &                                                          \\
7.05(0.03)     & 4$^{+}$ & 2  &  95(20)  & $\leq$120~\cite{bib:hah96}; 90$\pm$40~\cite{bib:har02}   \\
\hline \hline
\end{tabular}
%
\end{table}

\begin{table*}
\caption{\label{table2}Resonance parameters adopted in the calculation of the $^{14}$O($\alpha$,$p$)$^{17}$F reaction rate.}
\begin{tabular}{|c|c|c|c|c|c|c|c|}
\hline \hline
$E_x$ (MeV)$^a$ & $E_{res}$ (MeV)$^a$ & $J^{\pi}$ & $\Gamma_{\alpha}$ (eV) & $\Gamma_{p}$ (keV)   & $\Gamma_{p{\prime}}$ (keV) & $\Gamma$ (keV) & $\omega$$\gamma$ (MeV)\\
\hline
5.153$\pm$0.01 & 0.039  & 3$^{-}$  & 4.3$\times$10$^{-52}$$^a$         & 1.7$^a$          &                  & $\leq$15$^a$     & 3.0$\times$10$^{-57}$ \\
6.150$\pm$0.01 & 1.036  & 1$^{-}$  & 3.9$\pm$1.0$^b$                   & 37.8$\pm$1.9$^c$ & 15.9$\pm$0.7$^c$ & 53.7$\pm$2.0$^c$ & 1.2$\times$10$^{-5}$  \\
6.286$\pm$0.01 & 1.172  & 3$^{-}$  & 0.34$^a$                          & 20$\pm$15$^d$    &                  & 20$\pm$15$^d$    & 2.4$\times$10$^{-6}$  \\
7.05$\pm$0.03  & 1.936  & 4$^{+}$  & 22.6$\pm$3.2$^e$                  & 90$\pm$40$^f$    &                  & 90$\pm$40$^f$    & 2.0$\times$10$^{-4}$  \\
7.35$\pm$0.02  & 2.236  & 2$^{+}$  & 40$\pm$30$^f$                     & 70$\pm$60$^f$    &                  & 70$\pm$60$^f$    & 2.0$\times$10$^{-4}$  \\
7.62$\pm$0.02  & 2.506  & 1$^{-}$  & 1000$\pm$120$^f$                  & 72$\pm$20$^f$    & $<$2$^f$         & 75$\pm$20$^f$    & 3.0$\times$10$^{-3}$  \\
7.94$\pm$0.01  & 2.826  & 3$^{-}$  & (11$\pm$6.6)$\times$10$^{3}$$^g$  & 35$\pm$15$^g$    & 9.0$\pm$5.6$^g$  & 55$\pm$20$^g$    & 6.2$\times$10$^{-2}$  \\
8.11$\pm$0.01  & 2.996  & 3$^{-}$  & (6.3$\pm$3.9)$\times$10$^{3}$$^g$ & 20$\pm$4$^g$     & 4$\pm$3$^g$      & 30$^a$           & 3.5$\times$10$^{-2}$  \\
\hline \hline
\end{tabular}
\footnotesize \\
$^a$ From Hahn {\it et al.}~\cite{bib:hah96}; $^b$ From Fortune~\cite{bib:for12a}; $^c$ From Bardayan {\it et al.}~\cite{bib:bar12}; \\
$^d$ From present work; $^e$ From Fortune~\cite{bib:for12}; $^f$ From Harss {\it et al.}~\cite{bib:har02}; $^g$ From Almaraz-Calderon {\it et al.}~\cite{bib:alm12}.
\end{table*}

\begin{figure}
\includegraphics[width=8cm]{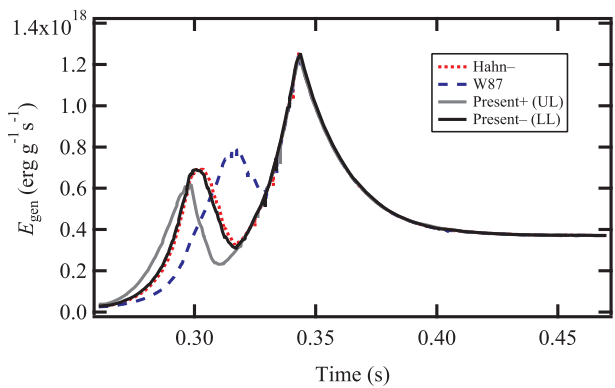}
\vspace{-3mm}
\caption{\label{fig5} (Color online) Nuclear energy generation rates during one-zone XRB calculations using the K04 thermodynamic history~\cite{bib:par08}.
Results using the ``Present" rates (black line for the lower limit (LL) of ``Present-", grey line for the upper limit (UL) of ``Present+") and the
``Hahn-" rates~\cite{bib:hah96} (red dotted line) are indicated. The result with Wiescher {\it et al.}~\cite{bib:wie87} rate is also shown for
comparison (labeled as ``W87"). See text for details.}
\end{figure}

The impact of the present new $^{14}$O($\alpha$,$p$)$^{17}$F rates has been examined using one-zone XRB models. With the representative K04
temperature-density-time
thermodynamic history ($T_\mathrm{peak}$=1.4 GK~\cite{bib:par08}), the nuclear energy generation rate ($E_\mathrm{gen}$) during an XRB has been
studied by performing separate post-processing calculations with different rates: the present rates (``Present+"~\&~``Present-" and their lower
and upper limits), as well as previous rates from Wiescher {\it et al.}~\cite{bib:wie87} (``W87"), Hahn {\it et al.}~\cite{bib:hah96}
(``Hahn+"~\&~``Hahn-"), Harss {\it et al.}~\cite{bib:har02}, and Alamaraz-Calderon {\it et al.}~\cite{bib:alm12}. Fig.~\ref{fig5} shows
$E_\mathrm{gen}$ at early times of the burst calculated using the upper limit of the ``Present+" rate and the lower limit of the ``Present-" rate.
$E_\mathrm{gen}$ curves calculated using the lower limit of the ``Present+" rate and the upper limit of the ``Present-" rate lie between these
two curves. The spread between the solid curves in Fig.~\ref{fig5} reflects the impact of the uncertainties of the present rates. Here only the
result with the ``Hahn-" rate is shown for comparison since the curve calculated using the ``Hahn+" rate is similar to that shown for the
``Present- (LL)" rate. Clearly, the sign of the interference has only a very marginal effect on the predicted $E_\mathrm{gen}$. The predicted
$E_\mathrm{gen}$ profiles using the rates of Harss {\it et al.} and Almaraz-Calderon {\it et al.} are not shown in Fig.~\ref{fig5} as these
profiles differ from that of the ``Present- (LL)" curve only between $\approx$0.30-0.32 s, where they lie between the two solid curves. The
$E_\mathrm{gen}$ profile calculated using the rate of Wiescher {\it et al.} (``W87") shows the largest differences from the $E_\mathrm{gen}$
calculated using the present rates. For example, at $\approx$0.31 s, $E_\mathrm{gen}$ calculated using the present rates is a factor of $\approx$3
less than that determined using ``W87".

Given the role of the $^{14}$O($\alpha$,$p$)$^{17}$F reaction in the breakout from the HCNO cycle during an XRB, it is not surprising that
different rates affect $E_\mathrm{gen}$ at early times. Nonetheless, the impact is decidedly modest: the largest shift in the initial peaks
observed for the different $E_\mathrm{gen}$ curves is only $\approx$0.01 s; this could be compared to the length of typical bursts
($\approx$10--100 s). Observing such a shift is certainly beyond the capabilities of current telescopes. As such, for the adopted model,
our results imply that the precise rate of this reaction has limited impact on the predicted nuclear energy generation during the burst.

Through a new $^{17}$F+$p$ resonant elastic scattering experiment, we have determined the energies, $J^\pi$ values and widths of three
$^{14}$O+$\alpha$ resonances between 6.1 and 6.4 MeV in $^{18}$Ne. We have firmly assigned $J^\pi$=1$^-$ to the 6.15 MeV state, resolving a
dispute in the literature. This state dominates the thermonuclear rate of the $^{14}$O($\alpha$,$p$)$^{17}$F reaction below $\sim$2 GK.
As well, we have found the evidence for a new state at $E_x$=6.85 MeV and discussed the possible structure origin. Using all available
experimental input, we have determined a new thermonuclear $^{14}$O($\alpha$,$p$)$^{17}$F rate and provided the quantitative tests
with an XRB model of the impact of multiple recent $^{14}$O($\alpha$,$p$)$^{17}$F rates. Contrary to many previous expectations in the
literature on the critical nature of this reaction rate, we find only minor variations in the predicted nuclear energy generation rates
when using reaction rates that differ by up to two orders of magnitude at the relevant temperatures. Indeed, the present rate and uncertainties
seem to be sufficient for calculations with the XRB model employed. Further tests using hydrodynamic XRB models are encouraged to confirm these results
and examine the impact of different $^{14}$O($\alpha$,$p$)$^{17}$F rates.

\begin{center}
\textbf{Acknowledgments}
\end{center}
We would like to thank the RIKEN and CNS staff for their kind operation of the AVF cyclotron. This work is financially supported by the NNSF
of China (Nos. 11135005, 11205212, 11321064), the 973 Program of China (2013CB834406), as well as supported by JSPS KAKENHI (No. 25800125). A.P was supported
by the Spanish MICINN (Nos. AYA2010-15685, EUI2009-04167), by the E.U. FEDER funds as well as by the ESF EUROCORES Program EuroGENESIS. A.A.C and D.I
were supported by the National Science and Engineering Research Council of Canada. K.I.H was supported by the NRF grant funded by the Korea
government (MSIP) (No. NRF-2012M7A1A2055625), and J.Y.M, H.S.J, and C.S.L by the Priority Centers Research Program in Korea (2009-0093817).

\end{document}